\documentstyle[prd,aps,epsf]{revtex}

\begin{document}

\twocolumn[\hsize\textwidth\columnwidth\hsize\csname
@twocolumnfalse\endcsname

\title{Isoperimetric inequality for higher-dimensional black holes}

\author{${}^{(a)}$Daisuke Ida and 
${}^{(b)}$Ken-ichi Nakao}

\address{${}^{(a)}$Department of Physics, Tokyo Institute of Technology, 
Tokyo 152-8550, Japan\\
${}^{(b)}$Department of Physics, Osaka City University, Osaka 558-8585, Japan}

\maketitle

\begin{abstract}
The initial data sets for the five-dimensional Einstein equation 
have been examined.
The system is designed such that the black hole ($\simeq S^3$) 
or the black ring ($\simeq S^2\times S^1$) can be found.
We have found that the typical length of the horizon can become
arbitrarily large but the area of characteristic closed two-dimensional
submanifold of the horizon is bounded above by the typical mass scale.
We conjecture that the isoperimetric inequality for black holes in $n$-dimensional
space is given by $V_{n-2}\lesssim GM$, where $V_{n-2}$ denotes the volume of typical 
closed $(n-2)$-section of the horizon and $M$ is typical mass scale, 
rather than $C\lesssim (GM)^{1/(n-2)}$ in terms of
the hoop length $C$, which holds only when $n=3$.
\end{abstract}
\vskip2pc
]

\vskip1cm

\section{Introduction}
There is much interest in higher dimensional space-times in the context of
the unified theory of elementary particles.
It is exciting if the existence of extra dimensions is confirmed in
high energy experiments.
In this aspect, the notion of the brane world \cite{brane} is an attractive idea.
This phenomelogical model provides us with a new way of thinking about our
universe, in which a size of the extra dimensions can be large because the
standard model particles and gauge interactions
are confined to the boundary of the higher-dimensional 
space-time. According to this scenario, 
the gravitational interaction at the short distance determined by the size of the
extra dimensions is modified effectively on the brane,
so that we might be able to see the extra dimensions by the gravitational experiments
below 1mm.
If the extra dimensions are large, the higher dimensional Planck scale may be given
by rather low energy. 
The possibility of TeV gravity, in which the fundamental Planck scale 
is set around TeV, has been much disscussed.

It is suggested that small black holes might be produced at LHC\cite{LHC1,LHC2}.
This argument follows from the hoop conjecture\cite{thorne}; a black hole with horizon forms
if and only if the typical length (hoop length) $C$ and the mass $M$ satisfies
$C\lesssim 4\pi G M$. Note that this statement might be valid only 
for four space-time dimensions.
The property of the higher-dimensional black holes have not so far been fully explored,
though there is much attention to this issue \cite{Tangherlini,kerr,ring,unique,n-hoop}.
We need reliable knowledge about such black holes to predict phenomelogical results.
We here consider black holes with small size compared with the extra dimensions, 
such that they are well described by the asymptotically flat black hole
solutions (treatment of Planck size black holes is beyond the scope of this paper).
The purpose of this paper is to consider the higher-dimensional 
generalization of the hoop conjecture.

In four dimensions, the hoop conjecture is believed to be valid.
Though it is loosely formulated, 
it seems to have at least following three meanings:
(i) If the massive object is compactified into a small region, there must be
a black hole;
(ii) A black hole is small;
(iii) Highly deformed black hole does not form.
The first one (i) has been proved by Schoen and Yau \cite{schoen-yau}
(see also Ref.~\cite{eardley}), which can be regarded
as the if part of the hoop conjecture.
A precise statement concerning the second proposition (ii) is for example given by
the Penrose inequality \cite{penrose-ineq,huisken,frauendiener}, 
which states that the square-root of the area of the apparent horizon $A$
is bounded above by the (ADM) mass: $\sqrt A\le 4\pi GM_{ADM}$.
Thus the Penrose inequality may serve as a part of the only if part of 
the hoop conjecture.
For the last statement (iii), which is also the only if part of the conjecture, 
we rely on the numerical works 
(e.g. Refs.~\cite{nakamura-shapiro,chiba,kt}). 
There is also a problem concerning the precise formulation of the conjecture,
such as the definition of the hoop length $C$.

At first glance, the hoop conjecture is not valid for higher-dimensional space-times,
since there is black string solutions.
In four dimensions, the length scale of the horizon cannot be so much larger than the
Schwarzschild radius, while this is not the case in higher dimensions.
The simplest example is the four-dimensional
Schwarzschild space-time times the real line, which is the five-dimensional vacuum solution
representing the gravitational field of the 
infinitely long $S^2\times {\bf R}$ black hole.

Nevertheless, we expect that higher dimensional black holes are also governed by
some isoperimetric inequality.
In what follows, 
we investigate initial data set for the five-dimensional Einstein equation and
estimate the size of the black holes.
Then we show the existence of such an 
isoperimetric inequality and give its physical reasoning.

\section{Momentarily Static Initial Data Set for the Five-Dimensional Einstein Equation}

Let us consider the initial data set $(g_{\mu\nu},K_{\mu\nu})$ on a four-dimensional
Cauchy surface $\Sigma^4$, where $g_{\mu\nu}$ is the induced metric on $\Sigma^4$
and $K_{\mu\nu}=g_\mu{}^\lambda{}^{5}\nabla_\lambda n_\nu$ 
 ($n_\nu$ denotes the unit normal to $\Sigma^4$) is the extrinsic curvature of
$\Sigma^4$.
The Hamiltonian and the momentum constraints are given by
\begin{equation}
R-K_{\mu\nu}K^{\mu\nu}+K^2=16\pi G\varrho
\label{Hamiltonian}
\end{equation}
and
\begin{equation}
\nabla_\nu(K^{\mu\nu}-Kg^{\mu\nu})=8\pi GJ^\mu,
\label{momentum}
\end{equation}
respectively, where $\varrho:={}^{5}G(n,n)$ denotes the energy density and
$J^\mu:=g^{\mu\nu}{}^{5}G_\nu(n)$ is the energy flux.
Let us consider the momentarily static initial data set
\begin{equation}
K_{\mu\nu}=0
\end{equation}
and assume the conformally flat metric
\begin{equation}
g=f^2\delta_{\mu\nu}dx^\mu dx^\nu,
\end{equation}
Then the momentum constraint~(\ref{momentum}) is solved with $J^\mu=0$ and
the Hamiltonian constraint~(\ref{Hamiltonian}) becomes
\begin{eqnarray}
\nabla_0^2 f
=-\frac{8\pi G}{3}f^3 \varrho, \label{eq:H-constraint}
\end{eqnarray}
where $\nabla_0$ denotes the flat connection.
We consider the vacuum case $\varrho=0$ so that an initial data is described by
a harmonic function $f$ on $E^4$.

We are interested in the possibility of the formation of highly non-spherical
black holes in higher dimensions.
As typical cases in five dimensions, we shall consider the initial data sets for
spindle, disk and ring shaped black holes.

\subsection{Spherical Black Holes}

 A spherically symmetric black hole gives reference values of the
volume, area and circumference of the horizon to the other cases 
discuessed below. 
We consider the metric with spherical symmetry of the form
\begin{equation}
g=f^2(r)\left[dr^2+r^{2}d\chi^2+r^{2}\sin^{2}\chi
(d\vartheta^2+\sin^2\vartheta^2d\varphi^2)\right],
\end{equation}
and consider a point source at the origin,
\begin{equation}
f^{3}\varrho={M_{ADM}\over 2\pi^{2}r^{2}}\delta(r),
\end{equation}
where $M_{ADM}$ is the ADM mass, that is, 
total gravitational mass of the system. 
Then the solution of Eq.~(\ref{eq:H-constraint}) becomes
\begin{equation}
f=1+{2GM_{ADM}\over 3\pi r^{2}}.
\end{equation}
This gives just an initial data for the Schwarzschild space-time.

The location of the black hole in the sense of the apparent horizon
is given by the minimal surface for the momentarily static initial data set $(K_{\mu\nu}=0)$
A spherical minimal surface centered at the origin satisfies 
\begin{equation}
\left(rf\right)_{,r}=0. 
\end{equation}
The solution of the above equation is given by
\begin{equation}
r=r_{\rm s}:=\left({2GM_{ADM}\over 3\pi}\right)^{1/2}.
\end{equation}
The volume of the minimal surface $H_{\rm s}$, area of its  
$S^{2}$-section $S_{\rm s}$ and the length of the circumference 
$C_{\rm s}$ of $S$ are given by
\begin{eqnarray}
{\rm Vol}(H_{\rm s})&=&2\pi^{2}\left[r_{\rm s}f(r_{\rm s})\right]^{3}
=2\pi^{2}\left({8GM_{ADM}\over 3\pi}\right)^{3/2}, \nonumber\\\\
{\rm Area}(S_{\rm s})&=&4\pi\left[r_{\rm s}f(r_{\rm s})\right]^{2}
={32GM_{ADM}\over 3}, \\
{\rm Length}(C_{\rm s})&=&2\pi r_{\rm s}f(r_{\rm s})
=2\pi\left({8GM_{ADM}\over 3\pi}\right)^{1/2},
\end{eqnarray}
respectively.

\subsection{Spindle Black Holes}
Let us consider the metric with axial and spherical symmetry of the form
\begin{equation}
g=f^2(\rho,z)\left[dz^2+d\rho^2+\rho^2(d\vartheta^2+\sin^2\vartheta^2d\varphi^2)\right],
\end{equation}
and consider the uniform line source of the length $L$ 
located at $z$-axis,
\begin{equation}
f^{3}\varrho={M_{ADM}\over 4\pi L\rho^{2}}\delta(\rho)
\theta\left(L/2-|z|\right),
\end{equation}
where $\theta$ is the Heaviside's step function. 

The solution of the Hamiltonian constraint~(\ref{eq:H-constraint}) is 
given by
\begin{eqnarray}
f&=&1+\frac{2GM_{ADM}}{3\pi L}\int_{-L/2}^{L/2}
\frac{dz'}{\rho^2+(z'-z)^2}\nonumber\\
&=&1+\frac{2GM_{ADM}}{3\pi L\rho}\left(
\arctan\frac{z+L/2}{\rho}-\arctan\frac{z-L/2}{\rho}\right).\nonumber\\
\end{eqnarray}
Note that this massive segment corresponds to another 
asymptotic end rather than
the singularity.
One may anyway fill up the segment with some spatially extended gravitational source.

\begin{figure}
\begin{center}
\leavevmode
\epsfxsize=0.45\textwidth
\epsfbox{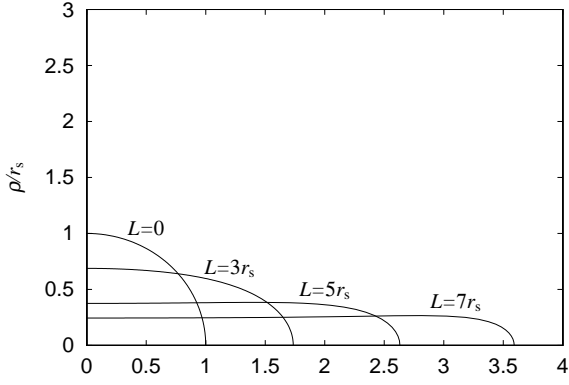}    
\caption{ 
Horizons for $L=0$, $3r_{\rm s}$, $5r_{\rm s}$ and $7r_{\rm s}$ 
are depicted in $(z,\rho)$-plane. 
The coordinate values are normalized by the radius $r_{\rm s}$ of 
the minimal surface of the spherical case $L=0$.
We will obtain minimal surfaces for $L>7r_{\rm s}$ 
if we wish. 
}
\label{fig:spindle-1}
\end{center}
\end{figure}

\begin{figure}
\begin{center}
\leavevmode
\epsfxsize=0.45\textwidth
\epsfbox{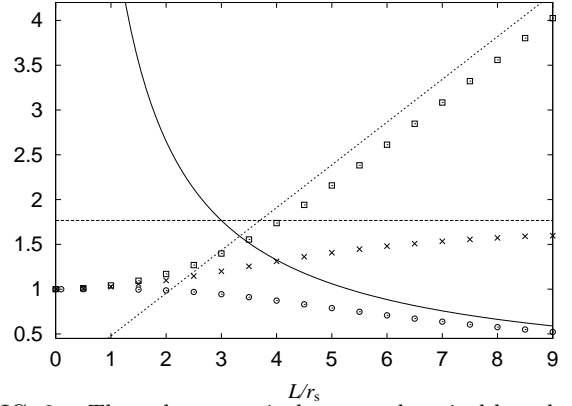}    
\caption{ 
The volume, typical area and typical length of a horizon 
are plotted as a function of the length $L/r_{\rm s}$.
These are normalized by those in the spherical case $L=0$:
Vol($H_{\rm s}$), Area($S_{\rm s}$) and Length($C_{\rm s}$), 
respectively.
The volume of the horizon $H$ is depicted by circles,  
area of the $S^{2}$-section $S$ 
($\vartheta=\pi/2$) 
of $H$ is depicted by crosses, and 
the length of the circumference $C$ of $S$ is depicted by a squares. 
The volume (solid line), area (dushed line) and length scale (dotted line) of 
a black string with the coordinate length $L$ are also plotted as 
a function of $L/r_{\rm s}$. 
This figure shows that the area is bounded above while 
the circumference is not. 
}
\label{fig:spindle-2}
\end{center}
\end{figure}

Due to the geometric symmetry imposed, we have only to consider 
the minimal surface equation for the three-surface:
$\rho=r(\xi)\sin\xi$, $z=r(\xi)\cos\xi$, given by
\begin{eqnarray}
&&r_{,\xi\xi}-4\frac{(r_{,\xi})^2}{r}-3r
+\frac{(r_{,\xi})^2+r^2}{r}
\left[
\frac{r_{,\xi}}{r}\cot\xi\right.
\nonumber\\
&&{}
-3\left(r_{,\xi}\sin\xi+r\cos\xi\right)
\frac{f_{,z}}{f}
\left.
+3\left(r_{,\xi}\cos\xi-r\sin\xi\right)
\frac{f_{,\rho}}{f}
\right]\nonumber\\
&&=0,
\end{eqnarray}
with the boundary condition:
$r_{,\xi}=0$ $(\xi=0,\pi/2)$, required from the regularity of the surface.
The results are shown in Figs.~\ref{fig:spindle-1} and \ref{fig:spindle-2}.

It seems that there always forms a black hole, however long
the massive segment is. 
When $L$ is sufficently large, the conformal factor 
near the origin behaves as that of the infinite line 
source,
\begin{equation}
f \sim 1+{2GM_{ADM}\over 3L\rho}.
\end{equation}
In the region where the conformal factor behaves as the above, 
the horizon is almost cylindrically symmetric and 
then is determined by
\begin{equation}
\left[\rho^{2}
\left(1+{2GM_{ADM}\over 3L\rho}\right)^{3}\right]_{,\rho}=0.
\end{equation}
The root of the above equation is given by
\begin{equation}
\rho=\rho_{\rm c}:={GM_{ADM}\over 3L}.
\end{equation}

The length scale of circumference $C_{\rm c}$ of $S_{\rm C}$,
the area of a section $S_{\rm c}$ ($\theta=\pi/2$),
and the volume of a cylindrical horizon $H_{\rm c}$
of the black string ($-L/2<z<L/2$) 
are given by
\begin{eqnarray}
{\rm Length}(C_{\rm c})&=&{3L\over 2\pi r_{\rm s}}{\rm Length}(C_{\rm s}), \\
{\rm Area}(S_{\rm c})&=&{9\pi\over 16}{\rm Area}(S_{\rm s}), \\
{\rm Vol}(H_{\rm c})&=&{27\pi r_{\rm s}\over 16L}{\rm Vol}(H_{\rm s}).
\end{eqnarray}
These quantities are also depicted in Fig.\ref{fig:spindle-2}. 
From this figure, we can see that length scale, area and  the volume
of a spindle black hole approach to corresponding black string 
values in the limit of $L\rightarrow\infty$.  

The area of the $S^2$-section $S$ 
of the horizon is always bounded above by the total mass:
\begin{equation}
\frac{{\rm Area}(S)}{32 GM_{ADM}/3}<O(1).
\label{a-bound}
\end{equation}

\subsection{Disk Black Holes}

 The result of the previous subsection shows that a horizon of
arbitrarily large linear size can form in five dimensions.
We here consider the possibility of disk shaped black holes.
The following metric is appropriate for this problem.
\begin{equation}
g=f^2(x,y)(dx^2+dy^2+x^2d\psi^2+y^2d\varphi^2).
\label{t2metric}
\end{equation}
This admits two orthogonal commuting Killing vector fields 
$\partial_\psi$, $\partial_\varphi$
and $\psi\sim\psi+2\pi$, $\varphi\sim\varphi+2\pi$ 
are regarded as the coordinates on $T^2$.

We consider a uniform massive disk as a source, 
\begin{equation}
f^{3}\varrho={M_{ADM}\over 2\pi^{2}D^{2} y}\delta(y)\theta(D-x).
\end{equation}
The gravitational field outside the above source 
is given by
\begin{eqnarray}
f&=&1+\frac{2GM_{ADM}}{3\pi^{2} D^2}\int_0^Dx'dx'\int_0^{2\pi}d\psi'\nonumber\\
&&\times
\frac{1}{(x-x'\cos\psi')^2+x'^2\sin^2\psi'+y^2}\nonumber\\
&=&1+\frac{2GM_{ADM}}{3\pi D^2}\ln\biggl|
{1\over 2y^{2}}\left[D^{2}-x^{2}+y^{2}\right. \nonumber \\
&&\left.+\sqrt{(x^{2}+y^{2})^{2}-2D^{2}(x^{2}-y^{2})+D^{4}}
\right]\biggr|.
\end{eqnarray}

Let us search for the apparent horizon of the form $x=r(\xi)\cos\xi$, $y=r(\xi)\sin\xi$,
determined by the differential equation
\begin{eqnarray}
&&r_{,\xi\xi}-4\frac{(r_{,\xi})^2}{r}-3r
+\frac{(r_{,\xi})^2+r^2}{r}
\left[
2\frac{r_{,\xi}}{r}\cot(2\xi)
\right.\nonumber\\
&&
\left.
{}-3\left(r_{,\xi}\sin\xi+r\cos\xi\right)
\frac{f_{,x}}{f}
+3\left(r_{,\xi}\cos\xi-r\sin\xi\right)
\frac{f_{,y}}{f}
\right]\nonumber\\&&=0,
\end{eqnarray}
subject to the boundary condition: $r_{,\xi}=0$ $(\xi=0,\pi/2)$.

\begin{figure}
\begin{center}
\leavevmode
\epsfxsize=0.45\textwidth
\epsfbox{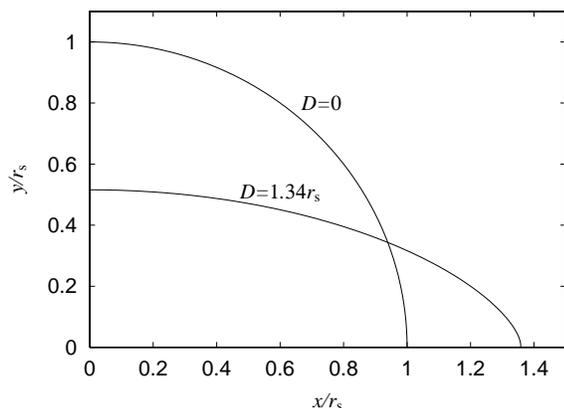}    
\caption{ 
Horizons of $D=0$ and $1.34r_{\rm s}$ 
are depicted in $(x,y)$-plane. 
The coordinate values are normalized by the radius $r_{\rm s}$ of 
the horizon in the spherical case $D=0$.
We could not find a horizon for $D>1.34r_{\rm s}$. 
}
\label{fig:disk-1}
\end{center}
\end{figure}

\begin{figure}
\begin{center}
\leavevmode
\epsfxsize=0.45\textwidth
\epsfbox{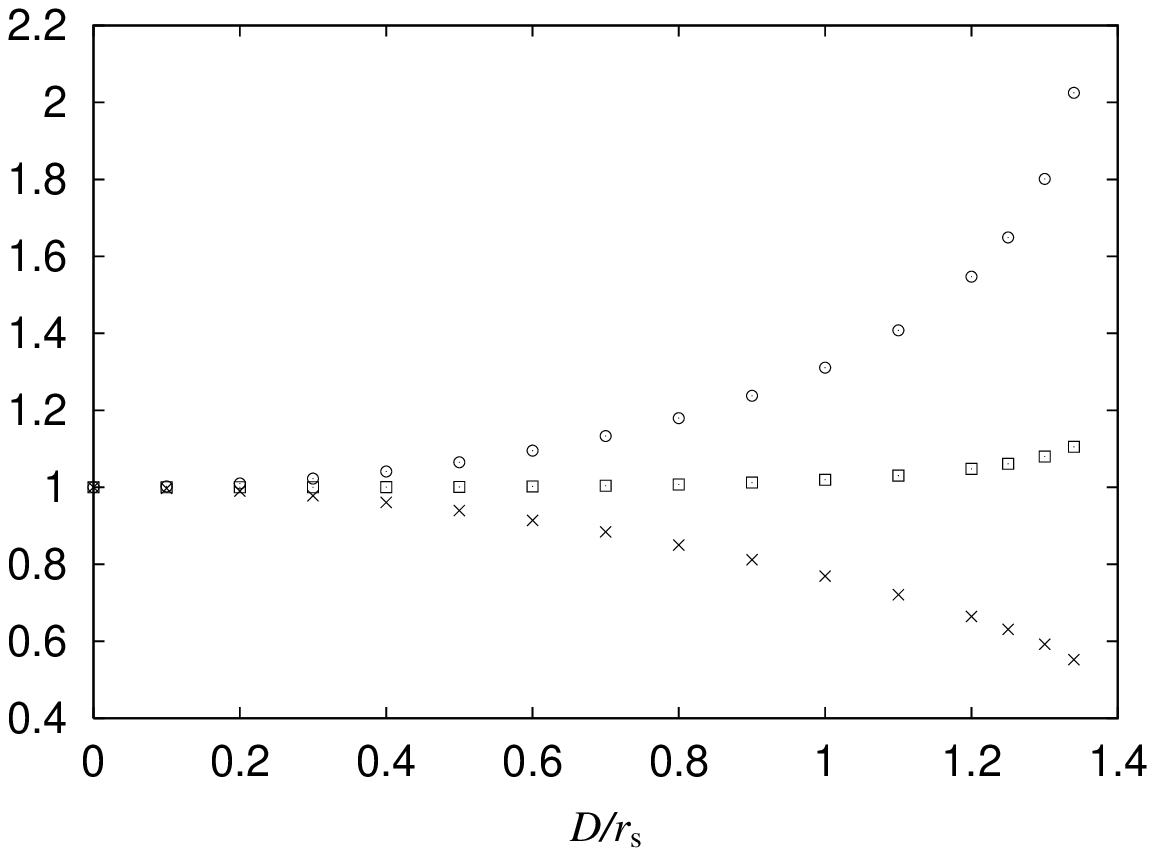}    
\caption{ 
The typical length scales 
${\rm Length}(C_{1})$ (squares), 
${\rm Length}(C_{2})$ (crosses) 
and ${\rm Length}(C_{3})$ (circles)
of a horizon 
are plotted as a function of the radius $D/r_{\rm s}$.
All quantities are normalized by those in the spherical case $D=0$.
}
\label{fig:disk-2}
\end{center}
\end{figure}

\begin{figure}
\begin{center}
\leavevmode
\epsfxsize=0.45\textwidth
\epsfbox{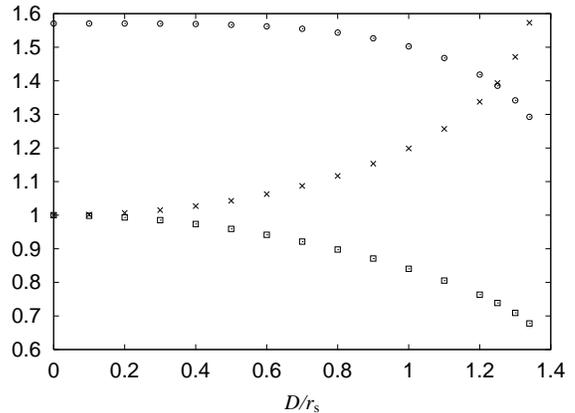}    
\caption{ 
The typical area scales Area($S_{1}$) (crosses), Area($S_{2}$) (squares) and 
Area($T$) (circles) of a horizon are plotted as a function of 
the disk radius $D/r_{\rm s}$. 
}
\label{fig:disk-3}
\end{center}
\end{figure}

\begin{figure}
\begin{center}
\leavevmode
\epsfxsize=0.45\textwidth
\epsfbox{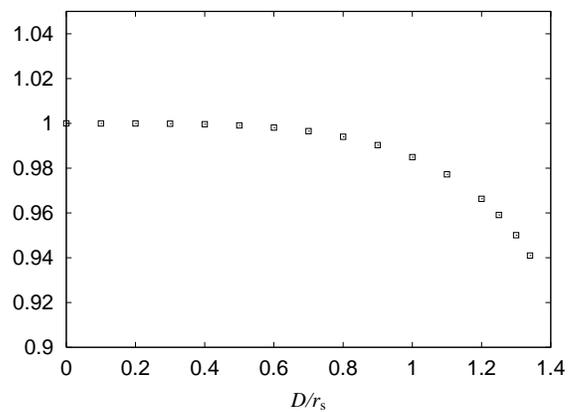}    
\caption{ 
The volume of a horizon is plotted as a function of 
the radius $D/r_{\rm s}$. 
The values are normalized by those in the spherical case $D=0$.
}
\label{fig:disk-4}
\end{center}
\end{figure}

The results are shown in Figs.\ref{fig:disk-1}$-$\ref{fig:disk-4}.
We evaluate typical hoop length scales
\begin{eqnarray}
{\rm Length}(C_1)&=&4\int_0^{\pi/2}f\sqrt{(r_{,\xi})^2+r^2}d\xi,\\
{\rm Length}(C_2)&=&{\rm max}\{2\pi fr\cos\xi;\xi\in[0,\pi/2]\},\\
{\rm Length}(C_3)&=&{\rm max}\{2\pi fr\sin\xi;\xi\in[0,\pi/2]\},
\end{eqnarray}
typical area scales
\begin{eqnarray}
{\rm Area}(S_{1})&=&4\pi\int_0^{\pi/2}f^2\sqrt{(r_{\xi})^2+r^2}r\sin\xi d\xi,\\
{\rm Area}(S_{2})&=&4\pi\int_0^{\pi/2}f^2\sqrt{(r_{\xi})^2+r^2}r\cos\xi d\xi,\\
{\rm Area}(T)&=&{\rm max}\{4\pi^2 f^2r^2\sin\xi\cos\xi;\xi\in[0,\pi/2]\},
\end{eqnarray}
and the volume of the horizon
\begin{equation}
{\rm Vol}(H)=2\pi^2\int_0^{\pi/2}r^{2}f^3\sqrt{(r_{,\xi})^2+r^2}\sin 2\xi d\xi.
\end{equation}

It can be seen that the inequality (\ref{a-bound}) still holds in this case.
Remarkably, a horizon does not form for large disks; we have not found 
a minimal surface for $D>1.34$.

\subsection{Black Rings}
In five-dimensional space-time, a black hole may have non-trivial topology \cite{topology},
while in four dimensions, the apparent horizon must be homeomorphic to sphere \cite{hawking}.
In particular, black rings homeomorphic to $S^2\times S^1$ are possible; Emparan and Reall
have found explicitly a stationary black ring solution \cite{ring}.
We here show the validity of the inequality (\ref{a-bound}) for this black ring case.
The metric used here is same as the disk case Eq.~(\ref{t2metric}).
The black ring will form if we put simply a uniform massive circle, 
\begin{equation}
f^{3}\varrho={M_{ADM}\over 4\pi^{2}Cy}\delta(x-C)\delta(y).
\end{equation} 

The conformal factor is then given by
\begin{eqnarray}
f&=&1+\frac{GM_{ADM}}{3\pi^{2}}\int_0^{2\pi}
\frac{d\psi'}{(x-C\cos\psi')^2+C^2\sin^2\psi'+y^2}\nonumber\\
&=&1+\frac{2GM_{ADM}}{3\pi\sqrt{(x+C)^2+y^2}\sqrt{(x-C)^2+y^2}}.
\end{eqnarray}
Let us search for the apparent horizon in the form:
$x=C+r(\xi)\cos\xi$, $y=r(\xi)\sin\xi$. This surface
is governed by the differential equation
\begin{eqnarray}
&&r_{,\xi\xi}-3\frac{(r_{,\xi})^2}{r}-2r
\nonumber\\
&&{}-\frac{(r_{,\xi})^2+r^2}{r}
\left[
\frac{r_{,\xi}\sin\xi+r\cos\xi}{r\cos\xi+C}
-\frac{r_{,\xi}}{r}\cot\xi
\right.\nonumber\\
&&
\left.
{}+3\left(r_{,\xi}\sin\xi+r\cos\xi\right)
\frac{f_{,x}}{f}
-3\left(r_{,\xi}\cos\xi-r\sin\xi\right)
\frac{f_{,y}}{f}
\right]\nonumber\\&&=0,
\end{eqnarray}
subject to the boundary condition $r_{,\xi}=0$ $(\xi=0,\pi)$.

\begin{figure}
\begin{center}
\leavevmode
\epsfxsize=0.45\textwidth
\epsfbox{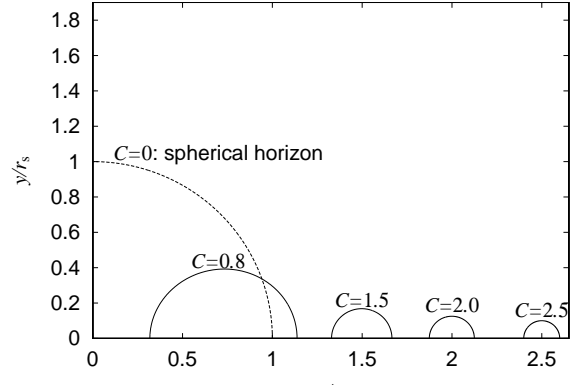}
\caption{
Black rings for $C=0.8 r_{\rm s}$, $1.5 r_{\rm s}$, $2.0 r_{\rm s}$, $2.5 r_{\rm s}$
are depicted in $(z,\rho)$-plane.
The coordinate values are normalized by $r_{\rm s}$.
A black ring can be found for $C=0.79r_{\rm s}$, but not for $C=0.78r_{\rm s}$.
}
\label{fig:ring-1}
\end{center}
\end{figure}

\begin{figure}
\begin{center}
\leavevmode
\epsfxsize=0.45\textwidth
\epsfbox{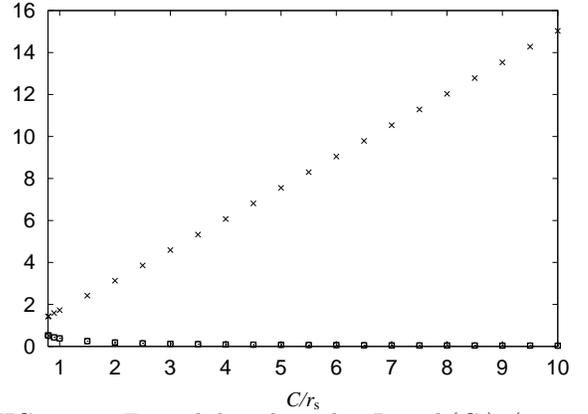}    
\caption{ 
Typical length scales ${\rm Length}(C_{1})$ (squares), 
${\rm Length}(C_{2})$ (crosses) and 
${\rm Length}(C_{3})$ (circles)
of black rings
are plotted as a function of the circle radius $C/r_{\rm s}$.
All quantities are normalized by those in the spherical case $C=0$.
}
\label{fig:ring-2}
\end{center}
 \end{figure}

\begin{figure}
\begin{center}
\leavevmode
\epsfxsize=0.45\textwidth
\epsfbox{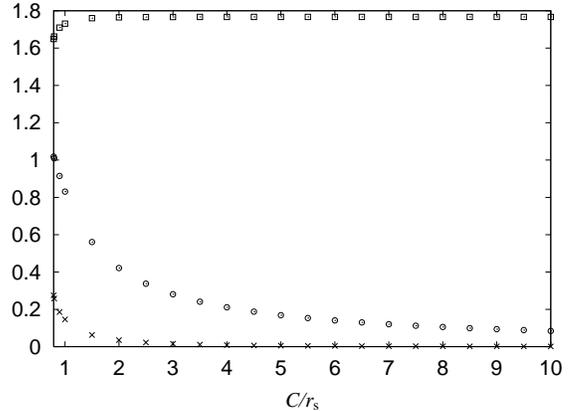}    
\caption{
The typical area scales Area($S$) (crosses), Area($T$) (squares) and 
the volume  Vol($H$) (circles) of a horizon are plotted as a function of 
the circle radius $C/r_{\rm s}$. 
}
\label{fig:ring-3}
\end{center}
\end{figure}

The results are shown in Figs.~(\ref{fig:ring-1}), (\ref{fig:ring-2}) and (\ref{fig:ring-3}).
We evaluate the typical hoop length scales
\begin{eqnarray}
{\rm Length}(C_1)&=&2\int_0^{\pi}f\sqrt{(r_{,\xi})^2+r^2}d\xi,\\
{\rm Length}(C_2)&=&{\rm max}\{2\pi fr\cos\xi;\xi\in[0,\pi]\},\\
{\rm Length}(C_3)&=&{\rm max}\{2\pi fr\sin\xi;\xi\in[0,\pi]\},
\end{eqnarray}
typical area scales
\begin{eqnarray}
{\rm Area}(S)&=&2\pi\int_0^{\pi}f^2\sqrt{(r_{\xi})^2+r^2}r\sin\xi d\xi,\\
{\rm Area}(T)&=&{\rm max}\{4\pi^2 f^2r^2\sin\xi\cos\xi;\xi\in[0,\pi]\},
\end{eqnarray}
and the volume of the black ring
\begin{equation}
{\rm Vol}(H)=4\pi^2\int_0^{\pi}f^3\sqrt{(r_{,\xi})^2+r^2}r\sin\xi
(r\cos\xi+C) d\xi.
\end{equation}


\begin{figure}
\begin{center}
\leavevmode
\epsfxsize=0.45\textwidth
\epsfbox{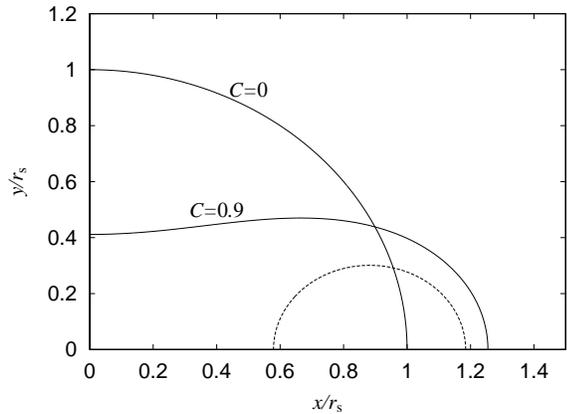}    
\caption{
Both a black hole and a black ring can be found for $0.79r_{\rm s}\le
C\le 0.90r_{\rm s}$. 
These horizons are depicted for $C=0.9 r_{\rm s}$.
A black hole with $C=0.91 r_{\rm s}$ cannot be found for a circle source.
}
\label{fig:hole-1}
\end{center}
\end{figure}

\begin{figure}
\begin{center}
\leavevmode
\epsfxsize=0.45\textwidth
\epsfbox{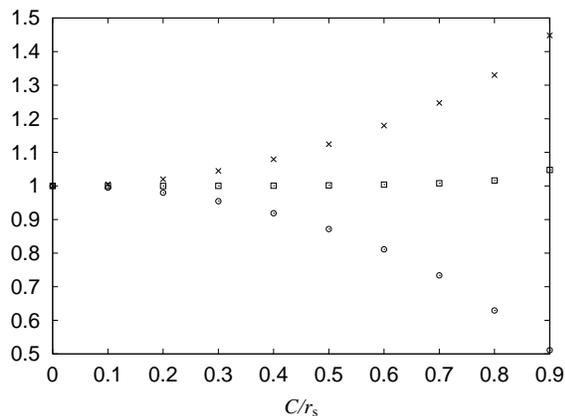}    
\caption{ 
The typical length scales 
${\rm Length}(C_{1})$ (squares), 
${\rm Length}(C_{2})$ (crosses) 
and ${\rm Length}(C_{3})$ (circles)
of a horizon 
are plotted as a function of the radius $C/r_{\rm s}$.
The definitions of these quantities are same as the disk case.
All quantities are normalized by those in the spherical case $C=0$.
}
\label{fig:hole-2}
\end{center}
\end{figure}

\begin{figure}
\begin{center}
\leavevmode
\epsfxsize=0.45\textwidth
\epsfbox{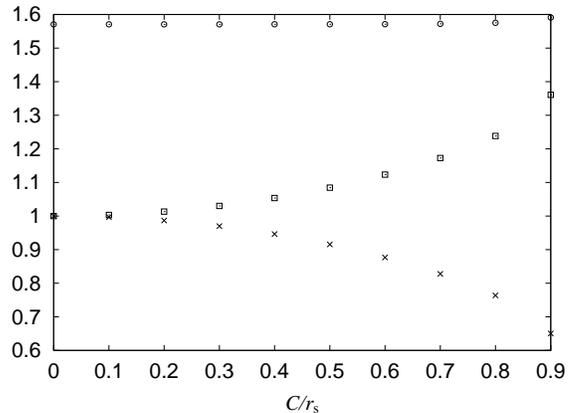}    
\caption{
The typical area scales Area($S_{1}$) (crosses), Area($S_{2}$) (squares) and 
Area($T$) (circles) of a horizon are plotted as a function of 
the circle radius $C/r_{\rm s}$. 
The definitions of these quantities are same as the disk case.
}
\label{fig:hole-3}
\end{center}
\end{figure}

\begin{figure}
\begin{center}
\leavevmode
\epsfxsize=0.45\textwidth
\epsfbox{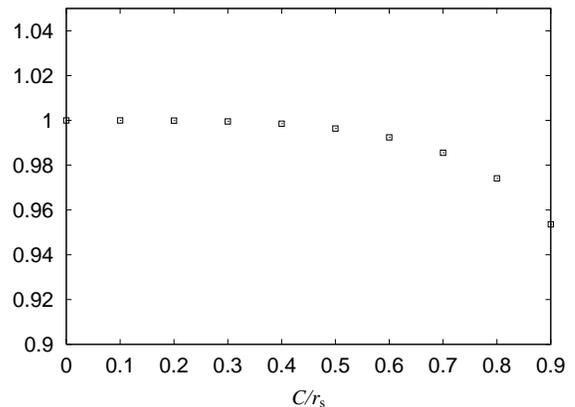}    
\caption{
The volume of a horizon is plotted as a function of 
the radius $C/r_{\rm s}$. 
The values are normalized by those in the spherical case $C=0$.
}
\label{fig:hole-4}
\end{center}
\end{figure}

For large radius of the massive circle, there always exists a black ring.
This can be expected from the result for the spindle case, since the local geometry
around a large circle resembles that around a line source.
On the other hand, a small circle makes a black hole homeomorphic to $S^3$
(See Figs.~\ref{fig:hole-1},~\ref{fig:hole-2},~\ref{fig:hole-3} and \ref{fig:hole-4}).
A new aspect found here is that both a black hole and a black ring form for a cirtain
range of the radius of the circle.
The inequality (\ref{a-bound}) is anyway satisfied, where the two-section of the horizon
can be characteristic sphere or torus.
\section{Conclusion}
We have investigated the momentarily static, conformally flat 
initial data sets for the five-dimensional Einstein equation.
We consider various configurations of the gravitational source and search for the apparent
horizons.

For the line source of the Euclidean length $L$, a black hole can be found for arbitrary $L$,
which can be contrasted with the corresponding four-dimensional situations.
In four dimensions, a black hole does not form 
when $L$ is much larger than the Schwarzschild radius.
The result here shows that the hoop length is not a good indicator of the horizon formation
in higher dimensions.
This can be interpreted as follows.
For the line source of the mass $M$ and the length $L\gg (GM)^{1/(n-2)}$
in $n$-dimensional space, 
the effective gravitational field at the symmetric hyperplane 
will have $(n-1)$-dimensional nature.
For the line source in four dimensional space-time, the effective gravity 
on the hyperplane will be that of
the $(2+1)$ dimensions, so that there does not form a black hole \cite{3D}.
In $(n+1)$-dimensional space-time, 
there will be a black hole of the radius $R\sim (GM/L)^{1/(n-3)}$.

From the above considerations, one can expect that the condition of horizon
formation is determined by the effective number of codimensions of 
the gravitational source. The horizon will not form if the effective number of
codimensions is less than three.
We have confirmed this expectation by studying the horizon formation due to the
disk source.
Since the effective gravity produced by the disk with the radius $D\gg (GM)^{1/2}$
is that of $(2+1)$-gravity, the horizon will not form for large disk sources.
In fact, the apparent horizon can be found only when $D$ is less
than or of order the Schwarzschild radius.
We found that the good indicator of the horizon formation is the typical area scale
of the system. In five-dimensional space-time, the condition for the horizon formation
will be given by the inequality
\begin{equation}
{\rm Area}\lesssim GM,\label{5-hoop}
\end{equation}
which can be regarded as the generalization of hoop conjecture for four-dimensional space-times.
In other words, the scale of typical codimension-two
submanifold of the horizon should be less than or of order the scale determined by the
mass scale. This argument is independent of the space-time dimensions. The corresponding
isoperimetric inequality for black holes in $(n+1)$-dimensional space-times will be
\begin{equation}
V_{n-2}\lesssim GM,
\end{equation}
where $V_{n-2}$ is the volume scale of characteristic codimension-two submanifold
of the horizon.

An interesting feature of higher-dimensional black holes is that the horizon can have
non-trivial topology.
In five-dimensional space-times, the horizon can be a black hole
($\simeq S^3$), a black ring ($\simeq S^2\times S^1$)
or their connected sums\cite{topology}.
For this reason, we have also investigated the condition for the black ring formation due to the
circle source. 
The inequality~(\ref{5-hoop}) still holds in this case. 
For large (small) circles, they form a black ring (hole). 
However, for appropriate ranges of the circle radius, both the black ring and the
black hole can be found such that the black hole encloses the black ring.
Thus we can expect that at the final stage of the gravitational collapse of the black ring,
new spherical black hole formes outside the black ring.

For large circle sources, the effective local gravity around the source will be that of
$(3+1)$ dimensions. This is the physical reasoning of the possibility of the black ring
in $(4+1)$ dimensions. 
While in $(4+1)$-dimensions, the torus horizon $(\simeq T^3)$ 
is forbidden. 
If all radiuses of $S^1$ of a torus source are large, then the effective gravity will be
$(2+1)$-dimensional, while if one of $S^1$ is small, then the horizon will not
be the torus anymore even if it forms.
Though there is no topology theorem for six or higher-dimensional black holes,
the torus topology might be forbidden.

For all examples studied here, the volume of the horizon is less than the 
spherical value:
\begin{equation}
{\rm Vol}(H)\le {\rm Vol}(H_{\rm s})=2\pi^{2}\left({8GM_{ADM}\over 3\pi}\right)^{3/2}.
\end{equation}
This inequality resembles the Penrose inequality for black holes in (3+1) dimensions.
Though the higher-dimensional generalization of 
the Penrose inequality is not known, similar inequality might exist.

\section*{Acknowledgments}
We would like to thank 
H.~Ishihara,
T.~Mishima,
K.~Nakamura, 
T.~Shiromizu, 
M.~Siino, 
M.~Yamaguchi, 
 for useful discussion
and comments.

\end{document}